\documentstyle[psfig,12pt]{article}
\author{Z. Kozio\l \thanks{electronic addresses: zkoziol@is.dal.ca;
http://is.dal.ca/$\sim$zkoziol/zkoziol.html}\space and R. A. Dunlap \\
{\it Department of Physics, Dalhousie University} \\
{\it  Halifax, N.S., Canada B3H 3J5}}
\title{{\bf Magnetostriction of a Superconductor:\\
-Results from the Critical-State Model}
}
\date{August 21, 1995}
\begin{document}
\maketitle
\begin{abstract}

In many cases, the critical-state theory can be treated as a
suffi ciently accurate approximation for the modelling of
the magnetic
properties of superconductors. In the present work, the
magnetostrictive
hysteresis is computed for a quite general case of the
modified
Kim-Anderson model.  The results obtained reproduce many
features of the
giant magnetostriction (butterfly-shaped curves) reported in
the
literature for measurements made on single-crystal samples
of  the
high-temperature superconductor $Bi_2Sr_2CaCu_2O_8$. It is
shown that addition
of a contribution to the magnetostriction in the
superconducting state
which is of similar origin as in the normal state, offers a
broader
phenomenological interpretation of the complex
magnetostriction hysteresis found in such heavy-fermion
compounds as $UPt_3$,
 $URu_2Si_2$ or $UBe_{13}$.
\end{abstract}

\vspace{5mm}
\newpage

   {\bf 1. Introduction.}

    \  Measurement of the magnetization has been the most
commonly used experimental technique for the study of flux-
pinning. Recently, however, magnetostriction studies have
been shown informative. For example, Ikuta et al. [1]
observed giant magnetostriction in $Bi_2Sr_2CaCu_2O_8$
single crystals in the superconducting state and showed that
the largest contribution to this effect comes from the
interaction between the critical currents induced by
changing the external field and the applied field, i.e., the
Lorentz force. Much attention has also been directed to
dilatometry studies of heavy-fermion superconductors. In
$UPt_3$ [2], the irreversible contribution to the
magnetostriction is small but in $UBe_{13}$  [3],
$UPd_2Al_3$ [4] and $URu_2Si_2$ [5], large sample size
changes are observed. \\
	In large fields, when the magnetic induction, $B$,
changes negligibly across the sample (e.g., in $URu_2Si_2$,
for $H\geq 10^3$ Oe [6]), the idealized situation for a slab
of thickness $2D$ may be written as $B(x)\simeq H-(4\pi
/c)j_c\cdot (D-x)$, with $x=0$ at the centre of the slab.
Here, $H$ is the external magnetic field and $j_c$ is the
critical current density, $(4\pi /c)j_c=-dB/dx$. The local
Lorentz force exerted on the current carriers,
$f=(1/c)(j_c\cdot B(x)$, is directed towards the centre of
the sample and leads to its compression, for increasing
external field. This force must be compensated by the
internal local stress, $\sigma (x)$; $\partial \sigma
/\partial x =(1/c)\cdot j_c \cdot B(x)$. Hence, with $j_c=-
(c/4\pi )\cdot B/dx$, we find that $\sigma (x)=-(H^2-
B^2(x))/8\pi $, for the boundary condition $\sigma (D)=0$.
For a material characterized by the elastic constant
$c_{11}$, the relative change of the sample size, $\Delta
D/D$, is given by:
\begin{equation}
\label{e1}
{\Delta D \over D} = {1 \over D}\cdot \int _0 ^D {\sigma (x)
\over c_{11}}dx ,
\end{equation}
which is equal to $-1/(8\pi c_{11} D)\cdot \int _0 ^D (H^2-
B^2(x))dx$. For the present case, with $(c/4\pi )\cdot
j_c\ll B/D$, we have: $(\Delta D/D\simeq -f_p /(2\cdot
c_{11}) \cdot D$. Hence, $\Delta D/D$ is directly related to
the flux-pinning-force density, $f_p=(1/c)\cdot B\cdot j_c$.

\vspace{5mm}
   {\bf 2. Magnetostriction in the modified Kim-Anderson
model}

   \ For qualitative calculations, we shall consider the
simplest possible geometry: a slab of thickness $2D$, with
the external magnetic field $H$ applied along its surface.
We use a modified Kim-Anderson formula for defining the
critical current density, $j_c(B)$, corresponding to a local
gradient of the magnetic induction:
\begin{equation}
\label{e2}
j_c = {\alpha \over (B+h)^n} = - (c/4\pi ) \cdot dB/dx .
\end{equation}
This dependence unifies many forms of the critical state
models. The calculating the magnetization for this geometry
and for $j_c(B)$ given by Eq. (2)  can be found in ref. [7].
This previous work [7] contains an analysis of other effects
observed in superconductors which can be explained using the
model of a critical-state. The present work provides
additional explanation of these phenomena and may be
considered as a continuation of our work on this subject.

The calculation of the magnetostriction consists of: {\it
i}) finding the magnetic field distribution  during the
magnetization process, {\it ii}) integrating the equation
$\partial \sigma / \partial x=  (1/c)\cdot j_c \cdot B(x)$
with the boundary condition that $\sigma (D)=0$ and {\it
iii}) using Eq. (1) for calculation of the magnetostriction.
The solution of the equation  (2), determines the flux
distribution and has the form: $B(x)=(\pm (4\pi /c)\cdot
\alpha \cdot (n+1)\cdot x+\beta )^{1/(n+1)} -h$, where the
integration constant $\beta$ is determined from the boundary
condition $B(\pm D)=H$, with the sign of $dB(x)/dx$
dependent on the most recent field change direction
occurring at point $x$. For the virgin magnetization curve,
$\beta$ is given by: $\beta =(H+h)^{n+1}-H_m$, where
$H_m=(H^*+h)^{n+1}-h^{n+1}$, and $H^*$ is the field of the
first full flux penetration into the centre of the slab,
$H^*= (\alpha \cdot (n+1)\cdot D+h^{n+1})^{1/(n+1)}$. Some
examples of the flux distribution for different situations
are drawn schematically in Fig. 1. In this figure, $B_0$ is
the magnetic induction in the sample centre and is given in
Table 1 for various stages of the magnetization process. The
magnetostriction during the virgin magnetization process,
for $H<H^*$, is found to be;
\vspace{3mm}
\begin{eqnarray}
\label{e3}
{8\pi c_{11} \cdot \Delta D(H) \over D}=H^2-{h^2 \over H_m}
 \cdot \left( (H+h)^{n+1}-h^{n+1}\right) \cr
+{(n+1) \over H_m} \cdot \left( {(H+h)^{n+3} -h^{n+3} \over
n+3 } -2h \cdot {(H+h)^{n+2} -h^{n+2} \over n+2} \right) .
\end{eqnarray}
In order to provide the results for the entire magnetization
loop in a transparent form, we define the following
quantities:
\vspace{3mm}

\centerline{\psfig{figure=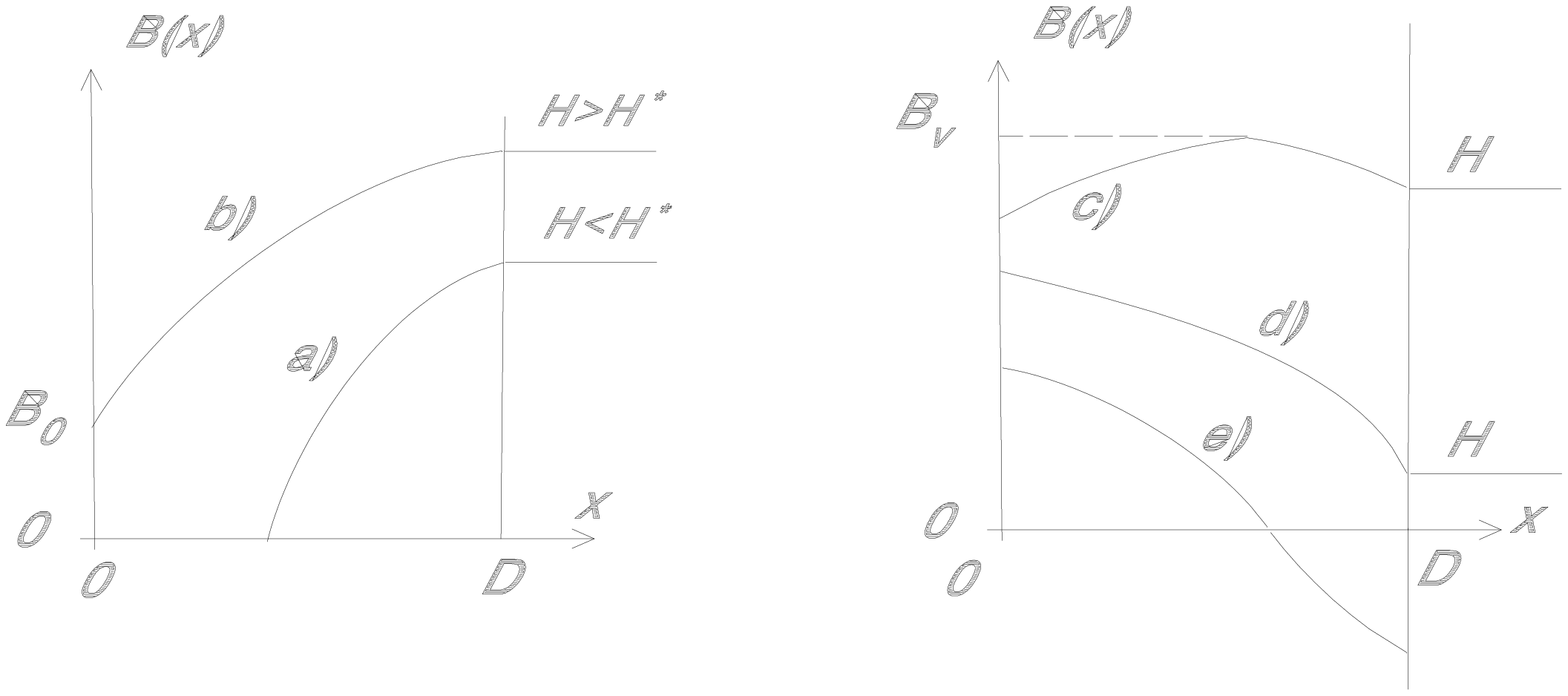,height=60mm,width=150mm,bbllx=0pt,bblly=0pt,bburx=350pt,bbury=750pt,angle=270}}
{\bf Figure 1}{\sl : The magnetic induction distribution in
a sample volume when {\bf a)} field increases from $0$ to
$H<H^*$, {\bf b)} field increases to $H>H^*$, {\bf c)} and
{\bf d)} when field decreases from $H_{max}$ to $H$, and
{\bf e)} when field decreases below $0$. $B_0$ is the field
at the slab centre. $B_v$ is the field at the kink of the
flux distribution at $x_v$, occurring for a certain range of
$H$ values after the sweep direction of $H$ is changed.
$x_v$  is equal to $1-((H_{max}+h)^{n+1}-
(H+h)^{n+1})/(2\cdot H_m)$.  $B_v$ and $B_0$  are  defined
in Table 1.}

\vspace{3mm}
{\bf Table 1.} Characteristic fields corresponding to the
curves a) to e) in

Fig.~1. $B_v$ is equal to $(0.5\cdot
((Hmax+h)^{n+1}+(H+h)^{n+1}))^{1/(n+1)}-h$.
\vspace{3mm}
{\footnotesize
\begin{tabular}{|l|l|l|l|l|}
\hline
Case & $H_{0,-}$ & $H_{0,+}$  & $H_{1,-}$  & $H_{1,+}$   \\
\hline
a)   & $0$       & $H$        & $0$        & $0$         \\
b)   & $B_0=((H+h)^{n+1} - H_m)^{1/(n+1)} -h$
               & $H$        & $0$        & $0$         \\
c)   & $B_0=((H_{max}+h)^{n+1} - H_m)^{1/(n+1)} -h$
               & $B_v$      & $B_v$      & $H$         \\
d)   & $B_0=((H+h)^{n+1} + H_m)^{1/(n+1)} -h$
               & $H$        & $0$        & $0$         \\
e)   & $B_0=(-(|H|+h)^{n+1}
+(H^*+h)^{n+1}+h^{n+1})^{1/(n+1)}$
               & $0$        & $0$        & $H$         \\
\hline
\end{tabular}
}
\vspace{3mm}

$F_{i,1}(H_{i,-},H_{i,+})=
h^2\cdot \left( (H_{i,+}+h)^{n+1}-(H_{i,-}+h)^{n+1} \right)
/H_m$,

$F_{i,2}(H_{i,-},H_{i,+})=
-(n+1)/(n+2)\cdot 2h \cdot \left((H_{i,+}+h)^{n+2}-(H_{i,-
}+h)^{n+2} \right) /H_m$,

$F_{i,3}(H_{i,-},H_{i,+})=
(n+1)/(n+3)\cdot \left((H_{i,+}+h)^{n+3}-(H_{i,-}+h)^{n+3}
\right) /H_m$,

with $H_{i,-}$ and $H_{i,+}$ defined in Table  1, for
situations corresponding to those in Fig. 1. The index $i$
enumerates different branches of the magnetic induction
curves of continuous $j_c(x)$ dependence. Using the above
equations, the magnetostriction can be written as
\begin{equation}
\label{e4}
{8\pi c_{11} \cdot \Delta D(H) \over D}=H^2-\sum_{i=1,2}
f(H_{i,-},H_{i,+}),
\end{equation}

where
\begin{eqnarray}
\label{e5}
&f(H_{i,-},H_{i,+})=S(H_{i,-},H_{i,+}) \cr
& \cdot \left( F_1(H_{i,-},H_{i,+}) +
F_2 (H_{i,-},H_{i,+}) +F_3 (H_{i,-},H_{i,+}) \right) .
\end{eqnarray}
The function $S(H_{i,-},H_{i,+})$ in Eq. (5) is equal to
$+1$ when $H_{i,-}\leq H_{i,+}$ and to $-1$, otherwise. The
remaining part of the magnetostriction hysteresis curve, not
covered in the Table 1, may be computed  using symmetry:
$\Delta D(-H)=\Delta D(H)$.

\vspace{3mm}
   {\bf 3. Discussion}

Some examples of the calculated magnetostriction hysteresis
curves are shown in Fig. 2. A large  variety of $\Delta
D(H)$ dependences is found: from $\Delta D(H)$  proportional
to $H$ at $H>H^*$, when $n=0$ (Bean model), to $\Delta D(H)$
strongly suppressed by field when $n>1$. For $n>0$, an
abrupt passing through $\Delta D(H)=0$ is found, when the
field sweep direction changes.

\centerline{
\psfig{figure=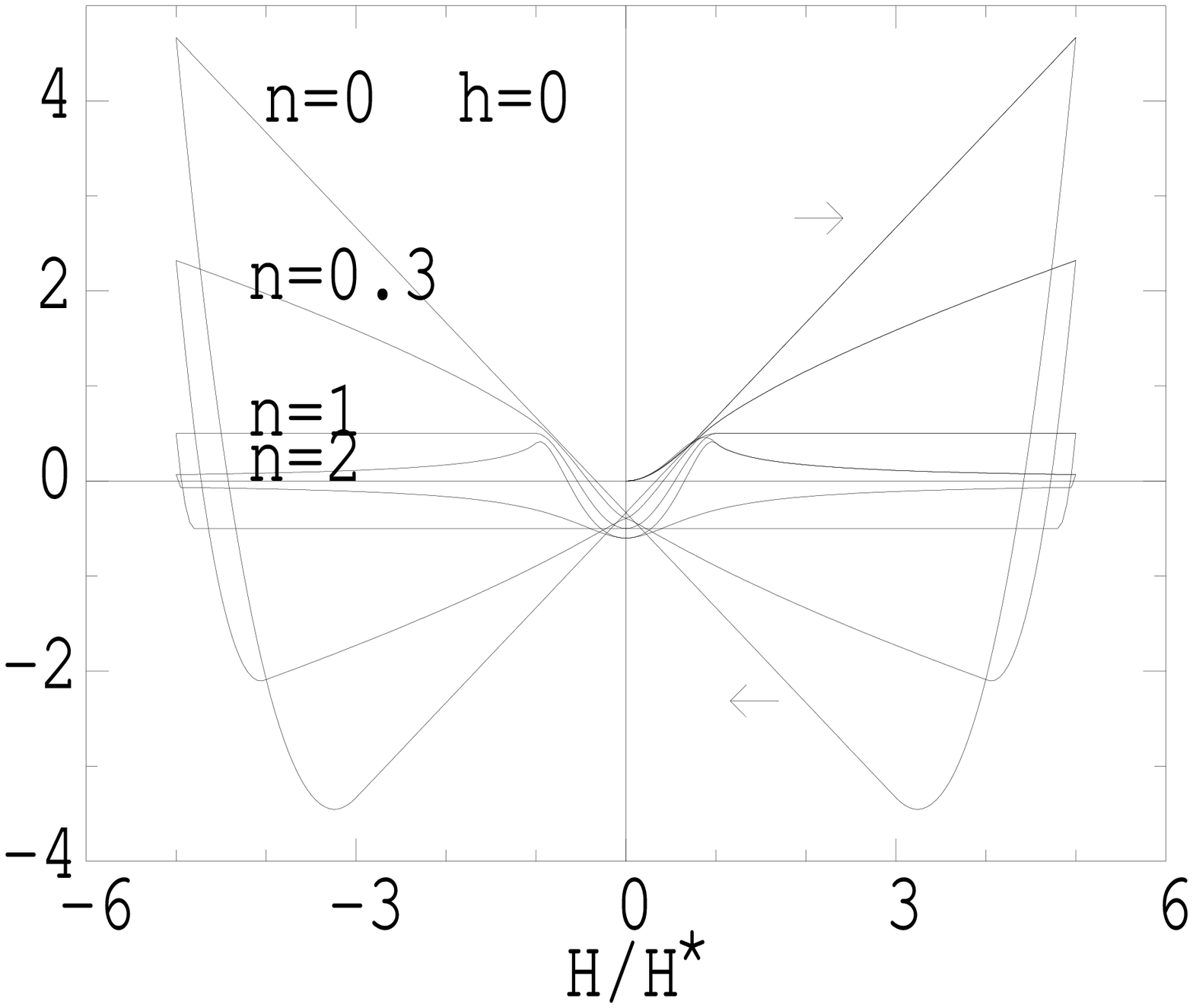,height=60mm,width=75mm,bbllx=0pt,bblly=0pt,bburx=600pt,bbury=700pt,angle=270}
\psfig{figure=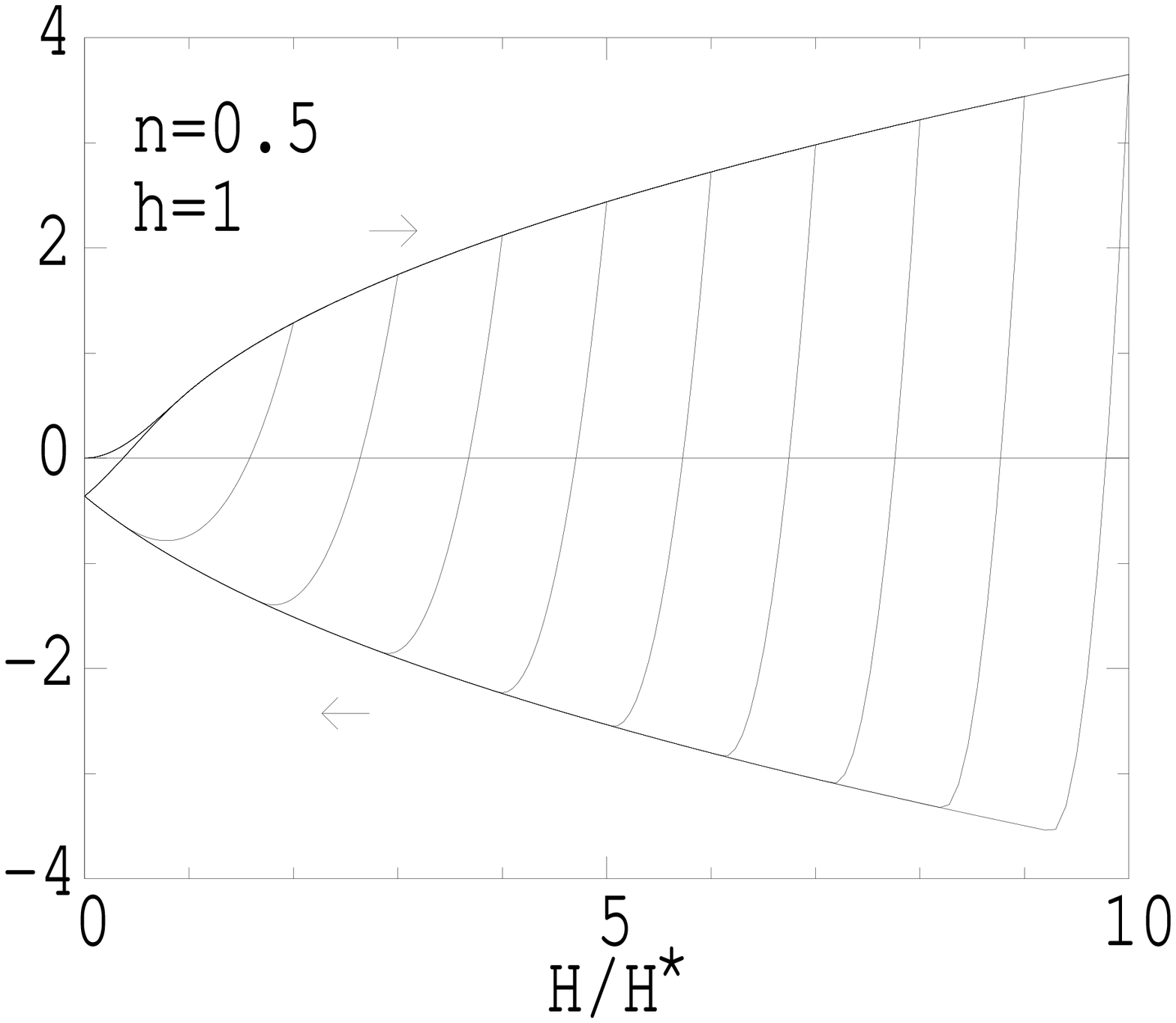,height=60mm,width=75mm,bbllx=0pt,bblly=0pt,bburx=600pt,bbury=700pt,angle=270}
}
{\bf Figure 2}{\sl : The calculated magnetostriction curves,
the virgin one and the entire hysteresis loop;  {\bf a)}
$h=0$, while $n$ changes from $0$ (the Bean model) to $n=2$,
{\bf b)} when $n=0.5$, $h=1$ while $H_{max}$ ranges between
$2$ and $10$. The vertical scale is in units of $8\pi \cdot
c_{11} \cdot \Delta D/D$. The arrows indicate the direction
of field change.}\\
The field-width of this transition region decreases strongly
with an increase of the maximum field applied (Fig. 2b).
This is due to a low value of the critical current density
at large fields and a decrease in the field change required
to reverse the direction of current flow in the sample
volume. A significant remanent magnetostriction due to a
frozen-in magnetic field is observed. The features discussed
above are commonly observed for high-$T_c$ and heavy-fermion
superconductors.

An unusual effect has been observed recently [5] for single-
crystal of $URu_2Si_2$. For fields near the upper critical
field, $H_{c2}(T)$, the irreversible contribution to  the
magnetostriction changes sign compared to the sign observed
at lower fields. Moreover, it was observed [5] that the sign
of the irreversible contribution to the magnetostriction for
some experimental configurations disagred with the
prediction of the critical-state model. On the other hand, a
large background contribution to the magnetostriction,
$\Delta D/D=b\cdot B^2$, with the parameter $b$ of the order
of $10^{-16}/G^2$, has been reported for heavy-fermion
materials [2,3,5], which is apparently of the same origin as
the normal-state magnetostriction. The role of this
contribution on the overall shape of the magnetostriction
may be considered. We note that the critical-state
magnetostriction component results from the calculation of
an integral over the $B^2(x)$ dependence. The total
magnetostriction in the superconducting state is then
assumed to be the sum of the critical-state and the normal-
state-like components. The sign of both contributions must
be considered carefuly; The critical-state magnetostriction
depends on the sign of the magnetic field and current while
the normal-state contribution depends on the sign of the
coefficient $b$. Using these arguments, we may write:
\begin{equation}
\label{e6}
{8\pi c_{11} \cdot \Delta D(H) / D}=H^2
- \sum_{i=1,2} \left( f(H_{i,-},H_{i,+})-G \cdot \left |
f(H_{i,-},H_{i,+}) \right | \right) ,
\end{equation}
with $f(H_{i,-}, H_{i,+})$ given by Eq. (5) and $G=8\pi
c_{11}\cdot b$. Magnetostriction calculated from Eq. (6) for
a few values of the parameter $G$ is shown in Fig. 3. An
important feature of these results is that both signs of the
magnetostriction and both signs of the irreversible
contribution to the magnetostriction hysteresis are
possible, while the shape of the curves might only be weakly
affected by the presence of a normal-state contribution
(except for the presence of a large background proportional
to $H^2$). The hysteresis of the normal-state-like
contribution to the magnetostriction in the superconducting
state is due to inhomogeneity of the magnetic induction
across the sample volume, which is caused by the presence of
critical current.    It is easy to show that at large
fields, in particular, near the upper critical field, both
contributions to the irreversible magnetostriction are
proportional to the critical current density. Hence, the
most intriguing property of the magnetostriction in some
heavy-fermion superconductors, i.e. the change of sign of
the irreversible part of the magnetostriction  as a function
of the applied field, can not be explained by this model, if
it is assumed that the parameter $G$ in Eq. (6) is
independant of the magnetic field value.

\centerline{\psfig{figure=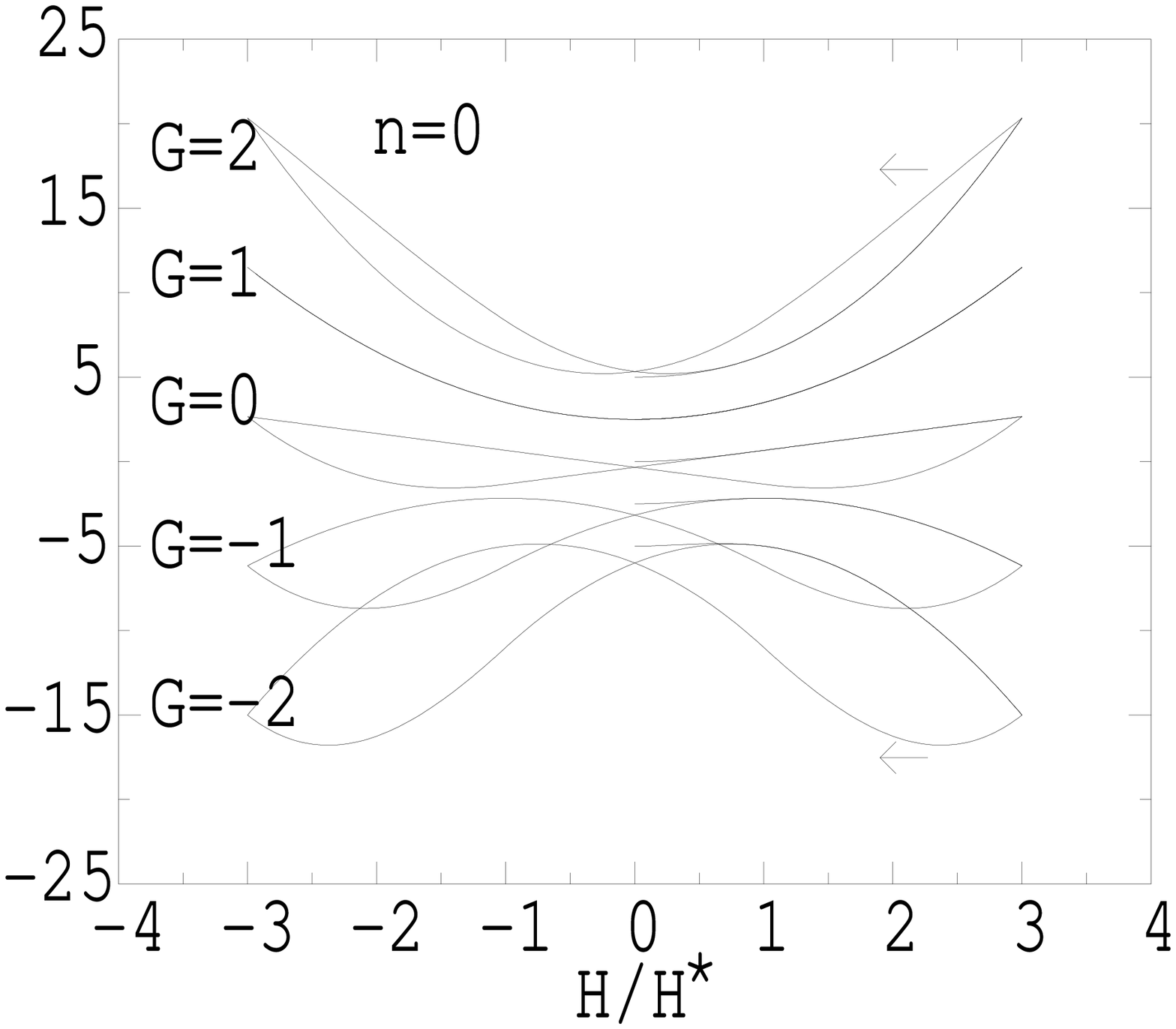,height=80mm,width=100mm,bbllx=0pt,bblly=0pt,bburx=600pt,bbury=650pt,angle=270}
}
{\bf Figure 3}{\sl : The calculated magnetostriction curves
for $n=0$ (the Bean model) for several values of the
parameter $G$, determining the normal-state contribution to
the magnetostriction: $G=2,1,0,-1$ and $-2$, for curves from
top to bottom, respectively. One must note that the values
of $G$ depend on the units of the magnetic field assumed,
which are here equal to $H^*$. The vertical scale is in
units of $8\pi \cdot c_{11} \cdot \Delta D/D$. It has been
shifted for some curves, for better clarity. The arrows
indicate the direction of field change.}\\

Using Eq. (5), magnetic measurements [6] in conjunction with
the value of the elastic constant $c_{11}$ allow for a
determination of $\Delta D/D$. In $URu_2Si_2$, $c_{11}=2.4
\cdot 10^{12} erg/cm^3$ [8], yields $2\Delta D/D=3.6\cdot
10^{-8}$ and $2\Delta D/D=6\cdot 10^{-8}$, for $H\|c$ and
$H\|a$, respectively, in reasonable agreement with measured
properties [5]. On the other hand, for $UPt_3$, the elastic
constant is larger, about $6\cdot 10^{12} erg/cm^3$ [9], and
the critical current density is much lower [7], resulting in
an irreversible contribution to the magnetostriction which
is hardly observable experimentally.

Gloos, et al. [4] reported significant difference between
the thermal-expansion curves measured during cooling in a
magnetic field through the superconducting transition and
those measured during heating with the same magnetic field
applied at low temperature, for the heavy-fermion
superconductor $UPd_2Al_3$. Any hysteresis of this kind
finds its phenomenological explanation within the critical
state model and is of the same nature as the magnetization
hysteresis in $M_{FC}(T)$ and in $M_{ZFC}(T)$ measurements,
as described by us in detail in ref. [7].

\vspace{5mm}
    {\bf 4. Conclusions}
     \ The present calculations have shown that the total
magnetostriction in the superconducting state can be
separated into critical-state and normal-state-like
components. In general, this approach has provided
predictions which are concistent with experimental results
in high-$T_c$ and heavy-fermion superconductors. In
particular, the model allows for the description of both
signs of the irreversible contribution to the
magnetostriction, though does not allow to explain the
mechanism causing the change of this sign as a function of
the magnetic field, which is observed in some heavy-fermion
systems.

\vspace{5mm}
{\bf Acknowledgements} This work has been made possible by a
Killam  Postdoctoral Fellowship awarded to Z.K. and by a
grant from the NSERC. The initial studies of the subject
have been carried out by Z.K. at University of Amsterdam.
These have been inspired by the experimental results (some
not published yet) of N. van Dijk, A. de Visser, V. Duijn,
and have been discussed with them as well with J.J.M. Franse
and P.F. de Ch\^atel.

\renewcommand{\refname}{\hspace{4cm} References}


\begin{thebibliography}{9}
\bibitem{ref1}H. Ikuta, et al., Phys. Rev. Lett. {\bf 70}
(1993) 2166.
\bibitem{ref2}N.H. van Dijk, et al, Physica  {\bf B186-188}
(1993) 267.
\bibitem{ref3}A. de Visser, et al.,  Phys. Rev. {\bf B45}
(1992) 2962.
\bibitem{ref4}K. Gloos, et al., Phys. Rev. Lett. {\bf 70}
(1993) 501.
\bibitem{ref5}N.H. van Dijk, et al.,  Phys. Rev. {\bf B51}
(1995) 12665.
\bibitem{ref6}Z. Kozio\l ,  et al. J. Magn. Magn. Mat. {\bf
140-144} (1994)  2065.
\bibitem{ref7}Z. Kozio\l ,  et al.,  Phys. Rev. {\bf B50}
(1994) 15978.
\bibitem{ref8}P. Thalmeier,  et al., Physica {\bf C175}
(1991) 61.
\bibitem{ref9}A. de Visser, Thesis, University of Amsterdam,
1986.
\end{thebibliography}
\end{document}